\documentclass[aps,pra,amsmath,amssymb,reprint,superscriptaddress]{revtex4-2}
\usepackage{physics}
\usepackage{graphicx}
\usepackage{dcolumn}
\usepackage{bm}
\usepackage{color}
\usepackage{url}
\usepackage[colorlinks]{hyperref}
\hypersetup{%
        plainpages=true,
        breaklinks=true,
        hypertexnames=false,
        pageanchor=true,
        colorlinks=true,
        linkcolor={blue},
        citecolor={magenta},
        urlcolor={blue},
        anchorcolor={black}
      } 
\usepackage{mleftright}
\newcommand{\figref}[1]{\mbox{Fig.~\ref{#1}}}

\newcommand{\secref}[1]{\mbox{Sec.~\ref{#1}}}

\renewcommand{\eqref}[1]{\mbox{Eq.~(\ref{#1})}}
\newcommand{\be}{\begin{equation}}
\newcommand{\ee}{\end{equation}}
\newcommand{\bea}{\begin{eqnarray}}
\newcommand{\eea}{\end{eqnarray}}
\begin{document}

\preprint{APS/123-QED}

\title{Global calibration of large-scale photonic integrated circuits}

\author{Jin-Hao Zheng}
\affiliation{CAS Key Laboratory of Quantum Information, University of Science and Technology of China, Hefei 230026, China}
\affiliation{School of the Gifted Young, University of Science and Technology of China, Hefei 230026, China}
\author{Qin-Qin Wang}
\affiliation{CAS Key Laboratory of Quantum Information, University of Science and Technology of China, Hefei 230026, China}
\affiliation{CAS Center for Excellence in Quantum Information and Quantum Physics, University of Science and Technology of China, Hefei 230026, China}
\author{Lan-Tian Feng}
\email{fenglt@ustc.edu.cn}
\affiliation{CAS Key Laboratory of Quantum Information, University of Science and Technology of China, Hefei 230026, China}
\affiliation{CAS Center for Excellence in Quantum Information and Quantum Physics, University of Science and Technology of China, Hefei 230026, China}
\affiliation{Hefei National Laboratory, University of Science and Technology of China, Hefei 230088, China}
\author{Yu-Yang Ding}
\affiliation{Hefei Guizhen Chip Technologies Co., Ltd., Hefei 230000, China}
\author{Xiao-Ye Xu}
\email{xuxiaoye@ustc.edu.cn}
\author{Xi-Feng Ren}
\author{Chuan-Feng Li}
\email{cfli@ustc.edu.cn}
\author{Guang-Can Guo}
\affiliation{CAS Key Laboratory of Quantum Information, University of Science and Technology of China, Hefei 230026, China}
\affiliation{CAS Center for Excellence in Quantum Information and Quantum Physics, University of Science and Technology of China, Hefei 230026, China}
\affiliation{Hefei National Laboratory, University of Science and Technology of China, Hefei 230088, China}

\date{\today}

\begin{abstract}
The growing maturity of photonic integrated circuit (PIC) fabrication technology enables the high integration of an increasing number of optical components onto a single chip.
With the incremental circuit complexity, the calibration of active phase shifters in a large-scale PIC becomes a crucially important issue.
The traditional one-by-one calibration techniques encounter significant hurdles with the propagation of calibration errors, and achieving the decoupling of all phase shifters for independent calibration is not straightforward.
To address this issue, we propose a global calibration approach for large-scale PIC.
Our method utilizes a custom network to simultaneously learn the nonlinear phase-current relations for all thermo-optic phase shifters on the PIC by minimizing the negative likelihood of the measurement datasets.
Moreover, the reflectivities of all static beams plitter components can also be synchronizedly extracted using this calibration method.
As an example, a quantum walk PIC with a circuit depth of 12 is calibrated, and a programmable discrete-time quantum walk is experimentally demonstrated.
These results will greatly benefit the applications of large-scale PICs in photonic quantum information processing.

\end{abstract}

\maketitle

\section{Introduction}
The diverse applications of photonic integrated circuits (PICs)\,\cite{Osgood2021} in the field of classical and quantum information processing\,\cite{Wang2020,Elshaari2020,Feng2020,Pelucchi2022,Feng:22,Moody2022,Giordani2023,Luo2023} have recently experienced significant growth, as they offer key attributes such as compactness, stability and scalability\,\cite{Su2023}.
In addition, PICs can be programmable to reconfigure the chip's functionality on demand, making them more versatile and flexible than application-specific static circuits\,\cite{Capmany2020}.
The programmability feature of PICs is typically achieved by actively controlling the tunable phase shifters.
Controlled phase shifts have been introduced in many ways\,\cite{Sun2022}, and the most commonly adopted approach is the thermo-optic effect\,\cite{Liu2022}, which utilizes electrically driven heaters to reversibly modify the refractive index of waveguides to impose phase shifts.
So far, a variety of experiments based on the programmable PICs have been reported\,\cite{Harris2017,Harris2018,Bogaerts2020,Bogaerts2020b,Giordani2023b}. 
However, as the increasing complexity of practical information processing tasks demands large-scale PICs, it becomes challenging to calibrate the significant number of active phase shifters on the chip.

The traditional calibration method requires the decoupling of all phase shifters followed by individual calibration for each one\,\cite{Shadbolt2012,Li2013,Miller2020,Alexiev2021,Cao2022}. 
However, this approach is not ideal for large-scale PICs due to two main challenges.
Firstly, while the traditional method can be used to calibrate the standard triangular ``Rech'' PICs\,\cite{Reck1994,Carolan2015}, rectangular ``Clements'' PICs\,\cite{Clements2016,Taballione2021,Pentangelo2024,Maring2024}, and PICs with cascaded phase shifters\,\cite{Qiang2018,Bao2023} by following a specific decoupling order, it is difficult to find an effective and direct way to decouple all phase shifters in the intricate mesh structures. 
Secondly, deviations in the calibration values of phase shifters in preceding levels may also spread errors in the calibration of subsequent phase shifters. 
As a result, calibration errors accumulate progressively with the increasing number of phase shifters, eventually becoming significant and non-negligible.

To address these issues, we present an efficient calibration method developed to globally calibrate all optical components on the PICs, thereby mitigating the accumulation of errors inherent in the traditional one-by-one calibration approach.
The global calibration method utilizes a custom network with simulation layers that creates a virtual duplicate of the physical PICs, modelling their practical nonunitary processes involving optical loss.
Using the gradient-descent approach, we can determine the optimal values of parameters that need to be calibrated, ensuring that the outputs of the network closely resemble the distributions of measurement data.
As an example, on a large-scale quantum walk (QW) PIC with a circuit depth of 12, we find that the trained custom network enables the simultaneous extraction of not only the nonlinear phase-current relations of all the thermo-optic phase shifters but also the reflectivities of all the beam splitters.
Moreover, we further experimentally demonstrate the silicon PIC calibration and implement the on-chip QW\,\cite{Grafe2016,Peruzzo2010,Sansoni2012,Crespi2013,Caruso2016,Tang2018,Qiang2021,wang2022large,He2024}.

This paper is organized as follows:
In \secref{sec2}, we will introduce our custom-architecture global calibration method, provide a concise overview of the QW PIC, and evaluate the performance of our calibration method using synthetic datasets featuring Gaussian noise.
In \secref{sec3}, we will showcase the calibration results of a silicon PIC utilizing noisy experimental datasets and discuss the performance regarding the experimental implementation of QW.
Finally, \secref{sec4} presents the summary of our work and engages in a discussion.

\section{\label{sec2}Theoretical Idea}

In this section, we will detail our method for achieving precise and fast calibration of PICs.
Our method is based on a custom network with simulation layers situated between the input and output layers, as depicted in Fig.\,\ref{fig:scheme}.
The entire custom network is trained using gradient-based optimization techniques to align its virtual process with the real physical process of the PICs, thereby allowing it to estimate the chip parameters that need to be calibrated.

\subsection{Global calibration method}

\begin{figure}[t]
  \includegraphics[width=0.45\textwidth]{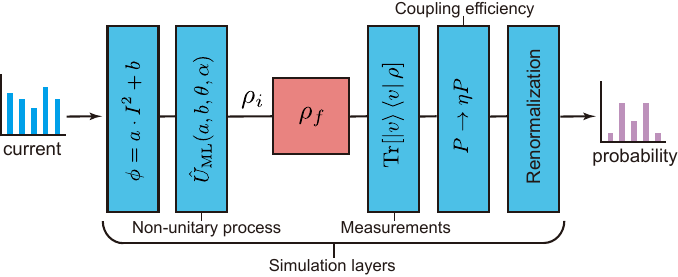}
  \caption{\label{fig:scheme} \textbf{Illustration of the custom architecture for global calibration.}
  The input data consists of the current distribution applied to the thermo-optic phase shifters.
  The first simulation layer of our custom network transforms the distribution of the current $\bm{I}$ into the distribution of the phase shift $\bm{\phi}$, with a subsequent layer outputting the parameterized final state $\rho_{f}$ for a given initial state $\rho_{i}$.
  The third simulation layer provides the corresponding measurement probability distribution of the final state $\text{Tr}[\left|v\right\rangle\!\!\left\langle v\right| \rho_{f}]$ in the computational basis $\ket{v}$.
  The fourth simulation layer introduces a parameter $\bm{\eta}$ into this probability distribution, which accounts for the relative fiber-chip coupling efficiency between output ports.
  The last layer renormalizes the probability distribution so that the sum of the probabilities over all output ports equals one, and a parameterized network distribution $\bm{P{_\text{ML}}}(\bm{\lambda})$ is generated thereafter.}
\end{figure}

PICs for linear optics comprise a cascade of passive beam splitters and active phase shifters arranged in various mesh structures\,\cite{Harris2018}.
The $2\times2$ unitary matrix for each on-chip beam splitter can be expressed as:
\be
\hat{U}_{\text{BS}}=\mqty(\cos\theta&\sin\theta\\-\sin\theta&\cos\theta),
\label{eq.bs}
\ee
where $\sin^2\theta$ denotes the reflectivity of the beam splitter.
Considering the inevitable optical loss in a practical chip, the transfer matrix for a lossy BS becomes nonunitary and is given by\,\cite{Uppu2016}:
\be
\hat{U}_{\text{LBS}}=\mqty(\sqrt{\alpha} &  0\\ 0 & \sqrt{\alpha})\mqty(\cos\theta&\sin\theta\\-\sin\theta&\cos\theta),
\label{eq.lbs}
\ee
where $\alpha$ denotes the transfer efficiency of a single BS, and 
$1-\alpha$ represents its optical loss.
The phase shifter provides a relative phase difference $\phi$ between the two output paths of the beam splitter, and its matrix representation is $\hat{U}_{\text{PS}} = \text{diag}\{1, e^{i\phi}\}$.
For the widely used thermo-optic phase shifter, the relative phase $\phi$ can be tuned by the current $I$ applied to the phase shifter, according to the following formula\,\cite{Liu2022}:
\begin{equation}\label{eq:5}
\phi= aI^2+b, 
\end{equation}
where $a$ and $b$ are hardware parameters to be determined.
Then, the nonunitary process $\hat{U}(\bm{a},\bm{b},\bm{\theta},\bm{\alpha})$ of PICs is the matrix product of its individual components.
Moreover, considering the mode-field diameter mismatch between the output ports of the PICs and the optical fiber arrays, it is necessary to introduce another parameter, $\bm{\eta}$, to indicate the fiber-chip coupling efficiency.
Consequently, $\bm{\lambda} = \{\bm{a},\bm{b},\bm{\theta},\bm{\alpha},\bm{\eta}\}$ are all undetermined parameters for a linear photonic circuit.

In contrast to the traditional one-by-one localized calibration method\,\cite{Shadbolt2012,Li2013,Miller2020,Alexiev2021,Cao2022}, we aim to globally calibrate all optical components on the PICs to reduce the accumulation of errors and avoid complex algorithms for decoupling the phase shifters.
To achieve this, we utilize a custom network with simulation layers that builds a virtual duplicate of the physical PICs, as illustrated in Fig.\,\ref{fig:scheme}.
The simulation layers of the network leverage prior knowledge of the structure of the PICs to be calibrated.
As a result, by utilizing the probability distributions of output ports, global calibration of the PICs could be implemented, thereby mitigating the high demand for complex quantum state and process tomography.

The learning process necessitates a number of training data, comprising randomly generated current distributions $\{\bm{I}\}_n$ applied to all the phase shifters and the target probability distributions $\{\bm{P}_{\text{tar}}\}_n$ of the output ports on the chip.
The index $n = 1, \cdots, N$ represents each training sample, and $N$ is the total number of training datasets.
We set the current distribution $\bm{I}$ as inputs, which are then fed to the input layer of our custom network.
The custom-network-parameterized nonunitary process $\hat{U}_{\text{ML}}(\bm{a},\bm{b},\bm{\theta},\bm{\alpha})$ can be determined using Eqs.\,(\ref{eq.lbs}) and (\ref{eq:5}).
For a given initial state $\rho_{i}$, the final state is given by $\rho_{f}=\hat{U}_{\text{ML}} \rho_{i} \hat{U}_{\text{ML}}^{\dagger}$, and the probability distribution $\text{Tr}[\left|v\right\rangle\!\!\left\langle v\right|\rho_{f}]$ can be obtained.
Considering the relative fiber-coupling efficiency $\bm{\eta}$ between the output ports of the PICs, the final probability distribution is normalized, and then the parameterized network distribution $\bm{P}_{\text{ML}}(\bm{\lambda})$ is generated.

We define the cost function $\mathcal{L}$ as the summation of the L1-norm distance\,\cite{Broome2010} between the target probability distributions in the training datasets $\{\bm{P}{_\text{tar}}\}_n$ and the parameterized network distributions $\{\bm{P_{\text{ML}}(\bm{\lambda})}\}_n$: $\mathcal{L}(\bm{\lambda}) = \sum_n\frac{1}{2} |\bm{P{_\text{ML}}}(\bm{\lambda})-\bm{P_{\text{tar}}}|$.
The aim of training the custom network is to find the optimal parameters $\bm{\lambda}^{\ast}$ by minimizing the cost function such that the parameterized network distributions approximate the target ones.
Then, the parameterized nonunitary process $\hat{U}_{\text{ML}}$ together with the relative coupling efficiency $\bm{\eta}^{\ast}$ for the trained network can well reflect the target physical process that generates the distributions $\{\bm{P_{\text{ML}}}(\bm{\lambda}^{\ast})\}_n\simeq\{\bm{P}{_\text{tar}}\}_n$.
%


\subsection{\label{sec2:A} Model of on-chip quantum walks}

\begin{figure}[t]
  \includegraphics[width=0.5\textwidth]{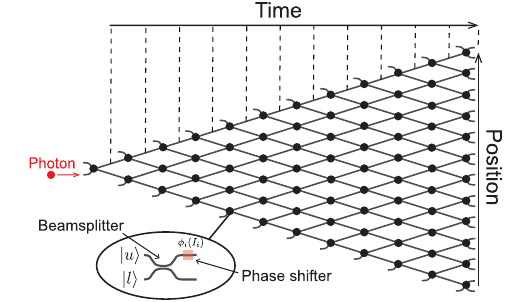}
  \caption{\textbf{Conceptual diagram of the QW PIC.} 
  Each black circle means a beam splitter element followed by a tuneable thermo-optic phase shifter (see inset), and black solid lines represent the waveguides. 
  At the last discrete time, the column of black circles represents only a beam splitter without a following phase shifter, as it does not affect the output probability distribution.
  \label{fig:chip}}
\end{figure}

Herein, we consider a QW PIC with a circuit depth of 12, illustrated in \figref{fig:chip}, as an example to validate our global calibration method.
We note that our method can also be applied to other mainstream types of PICs implementing universal unitary operations, such as the triangular ``Rech'' and rectangular ``Clements'' PICs\,\footnote{\label{Sup}See Supplemental Material at \href{http://link.aps.org/supplemental/10.1103/PhysRevApplied.22.054011}{http://link.aps.org/supplemental/10.1103/PhysRevApplied.22.054011} for the calibration results of the triangular ``Rech'' and rectangular ``Clements'' chips.}.
The QW PIC is calibration hard and shows difficulty in decoupling all phase shifters.
The whole physical process of the QW PIC can be mapped onto the discrete-time QW on a one-dimensional (1D) line\,\cite{Aharonov1993,Xu2018,Fenwick2024}.
The discrete-time evolution of a lossy QW can be described by the nonunitary process $\hat{U}=\prod_{t=1}^{T} (\hat S \hat M \hat C)$.
Here, $T$ denotes the number of discrete-time steps, $\hat{M}$ is the loss operator, $\hat S$ is the shift operator, which signifies the movement of the walker, and $\hat C$ is the coin operator, representing the coin-tossing process. 
In the tensor-product basis $\ket{v}=\ket{x}\otimes\ket{c}$, where $\ket{x}$ ($x \in \mathbb{Z}$) and $\ket{c}$ ($c=\downarrow,\uparrow$) express the position and coin space, respectively, $\hat C$, $\hat{M}$, and $\hat S$ can be written as\,\cite{Tao2021}:
\be
\hat C = \sum_{x}\ket{x}\bra{x} \otimes  \mqty(\cos\theta &  e^{i\gamma}\sin\theta\\ - e^{i\beta}\sin\theta & e^{i(\gamma+\beta)}\cos\theta),
\label{eq1}
\ee
\be
\hat M = \sum_{x}\ket{x}\bra{x} \otimes  \mqty(\sqrt{\alpha} &  0\\ 0 & \sqrt{\alpha}),
\label{eq.loss}
\ee
\be
\hat S = \sum_x (\ket{x-1}\bra{x}\otimes \ket{\downarrow}\bra{\downarrow}+\ket{x+1}\bra{x}\otimes \ket{\uparrow}\bra{\uparrow}).
\ee
Here, the parameter $\alpha$ determines the loss strength and varies depending on both the position and time step.
$\cos^2\theta$ and $\sin^2\theta$ denote the probabilities of a walker moving backwards and forwards on the 1D line, respectively. 
Parameters $\beta$ and $\gamma$ introduce the relative phase difference. 
The coin operator $\hat C$ can be position and time dependent, a feature necessary for specific tasks in QW-based quantum simulations\,\cite{Crespi2013} and quantum algorithms\,\cite{Harris2017}.

For the QW PIC, the coin state $\{\ket{\uparrow},\ket{\downarrow}\}$ in the QW model can be cast on the path state (inset of Fig.\,\ref{fig:chip}): the upper path of the beam splitter corresponds to $\ket{u}$, while the lower path corresponds to $\ket{l}$.
The localized coin state at each position is described by $\ket{c}=c_1\ket{u}+c_2\ket{l}$, which can be interpreted as a superposition state located on the two paths of the corresponding beam splitter. 
Each beam splitter with a splitting ratio of 50:50 and phase shifter work together (black circle in Fig.\,\ref{fig:chip}) to realize the coin operator $\hat{C}$ in the sense of path states, corresponding to Eq.\,(\ref{eq1}) with $\theta=\pi/4$ and $\beta=0$.
The interwoven mesh architecture of the PIC, composed of the on-chip waveguides (black solid lines), realizes the shift operators $\hat{S}$. 
After $T$-step nonunitary dynamics of the QW PIC, the size of the position space is $T$, with each position having two paths representing the coin space. 
Therefore, the total number of output ports of the QW PIC is $2T$.

\subsection{\label{sec2:C} Numerical benchmarks}

\begin{figure*}
  \centering
  \includegraphics[width=1.0\textwidth]{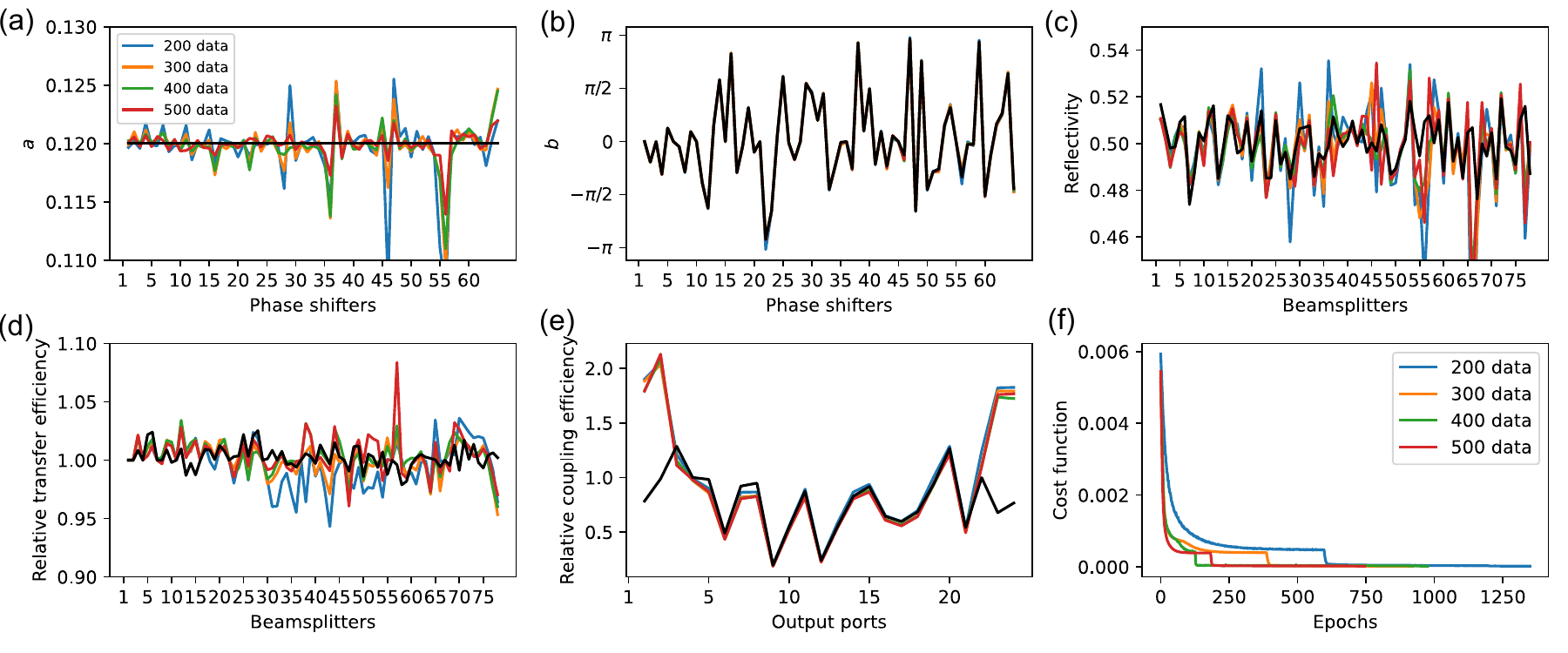}
  \caption{\textbf{Calibration results using synthetic datasets.} %
  (a)$\sim$(e) exhibit the calibration values of $\bm{a}$, $\bm{b}$, $\sin^2\bm{\theta}$, $\bm{\alpha}$, and $\bm{\eta}$, respectively, given different numbers of training samples.
  Black solid lines in (a-e) are the preset target values in theory.
  The indices of the phase shifters and beam splitters on the QW PIC are arranged in order from the first column on the far left to the last column, with each column arranged from top to bottom.
  The indices of the output ports are ordered from top to bottom.
  The significant deviations in the relative coupling efficiencies $\bm{\eta}$ of the edge output ports at the top and bottom positions are attributed to the extremely low output probability (nearly $10^{-4}$) at these four ports.
  (f) The values of the cost function versus the number of epochs.
  }\label{fig:sim_res}
\end{figure*}

\begin{figure*}
  \centering
  \includegraphics[width=1.0\textwidth]{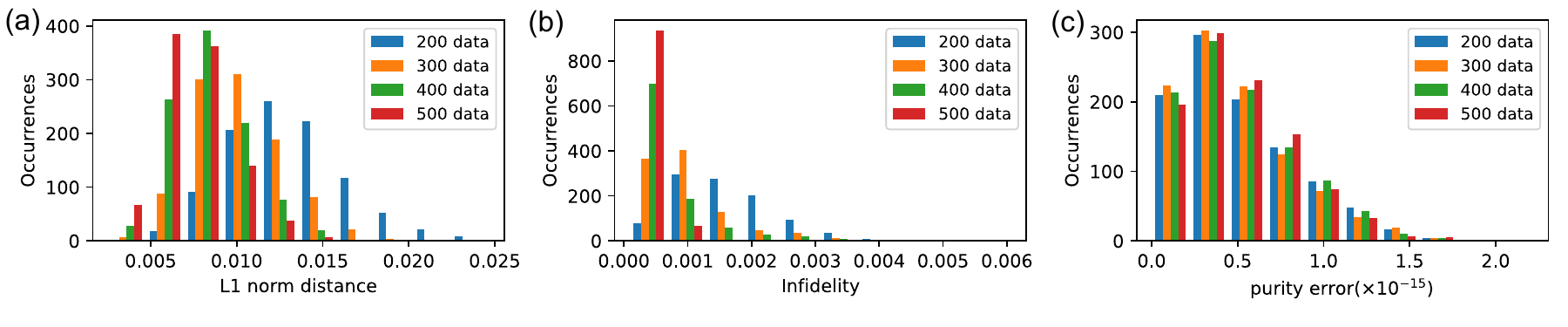}
  \caption{\textbf{Benchmarking the performance of the global calibration method.} Histogram of the values of (a) L1-norm distance, (b) infidelity and (c) purity error for 1000 test samples, given different numbers of training samples.
  }\label{fig:4}
\end{figure*}

Before training the custom network with noisy experimental results, we begin by numerically benchmarking the global calibration method using synthetic datasets.
To effectively simulate noisy experimental environments and assess the robustness of our calibration method against external noise, we introduce Gaussian noise into the data generation process for the synthetic datasets\,\cite{Ahmed2023}.
Here, the Gaussian noise is added to the output probabilities to mimic experimental measurement errors, such as small vibrations of the optical fibers coupled to the output ports of the PICs.
The noise is sampled from a zero-mean Gaussian distribution $\mathcal{N}(0,\epsilon)$ with a standard deviation of $\epsilon=0.01$ defining the noise degree.
Then, the calibration process aims to minimize the cost function $\mathcal{L}(\bm{\lambda}) = \sum_n \frac{1}{2}|\bm{P}{_\text{ML}}(\bm{\lambda})-\bm{P}{_\text{th}}|$ for given datasets of noisy synthetic probability distributions $\{\bm{P}_{\text{th}}\}_n$.

The QW PIC comprises 66 phase shifters, 78 beam splitters, and 24 input-output ports.
The initial state is fixed to be a localized state $\rho_i=\left|\psi_{\text{i}}\right\rangle\!\!\left\langle \psi_{\text{i}}\right|$ with $\ket{\psi_{\text{i}}} = \ket{x_0}\otimes\ket{l}$, corresponding to the lower path of the first waveguide.
The target values for the reflectivity of the BSs $\sin^2\bm{\theta}$, the relative transfer efficiency $\bm{\alpha}$ between BSs, and the relative output coupling efficiency $\bm{\eta}$ between output ports are set randomly near their ideal values of 0.5, 1, and 1, respectively.
The target values for the coefficient $\bm{a}$ are set to an empirical value of 0.12, while the values for $\bm{b}$ are randomly selected from the interval $[-\pi,\pi]$.
The cost function $\mathcal{L}(\bm{\lambda})$ is minimized with Adam optimizer that updates the network parameters $\bm{\lambda}=\{\bm{a},\bm{b},\bm{\theta},\bm{\alpha},\bm{\eta}\}$ for each epoch.
The Adam optimizer is configured as follows\,\cite{kingma2017adam}: the initial learning rate is 0.01 and decreases by multiplication with a factor of 0.99 after each epoch.
All other core parameters of Adam are maintained at their default values.

\begin{figure*}
  \centering
  \includegraphics[width=1.0\textwidth]{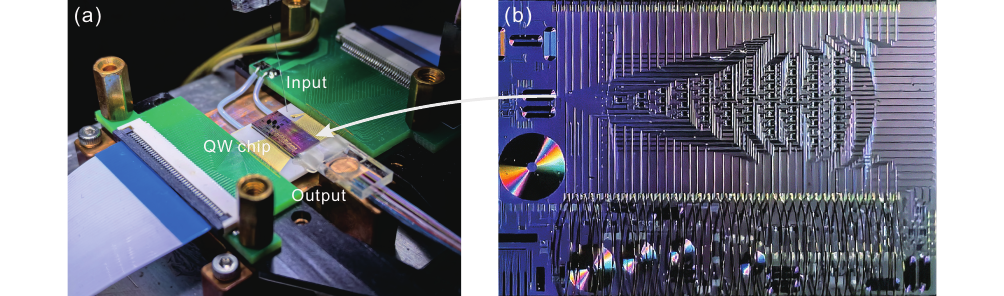}
  \caption{\textbf{Experimental on-chip QWs and external setup.} 
  (a) The experimental device includes a laser input connected to the photonic QW chip and an array of optical fibers coupled to the output ports.
  (b) A close-up view of the chip depicted in (a). The chip consists of 66 phase shifters, 78 beam splitters, 24 input ports, and 24 output ports.}\label{fig.setup}
\end{figure*}

During the iteration process, simultaneously updating all network parameters using the gradient-descent method is susceptible to getting trapped in a local minimum of $\mathcal{L}$.
To mitigate this issue and escape local minima, we utilize a nonsimultaneous parameter update method, updating only partial parameters at each epoch while keeping the remaining parameters fixed.
The relative coupling efficiency $\bm{\eta}$ is configured to be constantly updated.
The reflectivity parameter $\bm{\theta}$ and the relative transfer efficiency $\bm{\alpha}$ are frozen at first.
We then alternatively train one of the two parameters $\{\bm{a},\bm{b}\}$ while the remaining one is kept frozen and untrainable.
Within a single alternating training process, the training for each parameter ceases upon meeting the cut-off condition, where five consecutive decreases in the loss function between adjacent epochs are each less than $10^{-5}$.
After eight rounds of alternating training, we proceed to train the last two parameters, $\bm{\theta}$ and $\bm{\alpha}$, while maintaining the fixed values of $\bm{a}$ and $\bm{b}$.
In the end, following the completion of the nonsimultaneous parameter updates described above, we proceed to simultaneously update all parameters $\bm{\lambda}$ until the cut-off condition is met.

We examine the impact of varying training sample numbers $N$ on the performance of our calibration method, as depicted in Fig.\,\ref{fig:sim_res}.
The optimal parameters $\bm{\lambda}^{\ast}$ of the trained custom network are illustrated in Figs.\,\ref{fig:sim_res}(a)-\ref{fig:sim_res}(e) for $N=200$\,(blue), 300\,(orange), 400\,(green), and 500\,(red).
It is observed that as the number of training samples increases, the values of the optimal parameters approach the preset target ones.
The corresponding training process of the custom network is displayed in Fig.\,\ref{fig:sim_res}(f), where the cost function is plotted as a function of the training epochs.
Compared to the slow-descent regions, each distinctly sharp drop in the cost function occurs as a result of one round of the alternating training of $\bm{a}$ and $\bm{b}$.

To quantify the performance of the calibrated PIC, we first compute the L1-norm distance $d(\bm{\lambda}^{\ast}) = \frac{1}{2} |\bm{P_{\text{ML}}}(\bm{\lambda}^{\ast})-\bm{P_{\text{tar}}}|$ between the target distributions and the network's predictions for 1000 test samples.
Furthermore, to fully characterize the output states, we also provide infidelity $F_{\text{in}} = 1-\text{Tr}(\sqrt{\sqrt{\rho_{\text{tar}}} \rho_{\text{ML}}(\bm{\lambda}^{\ast}) \sqrt{\rho_{\text{tar}}} })$ as a measurement of the distance between the target and network-predicated output quantum states.
The values of the L1-norm distance and infidelity range from 0, indicating a perfect match, to 1, indicating a complete mismatch.
Figs.\,4(a)-4(c) show the histograms of the L1-norm distance, infidelity and purity error $|\text{Tr}[\rho^{2}_{\text{ML}}(\bm{\lambda}^{\ast})] - \text{Tr}[\rho^{2}_{\text{tar}}]|$ for the test datasets, using different numbers of training samples $N$, respectively.
We can see that nearly all the values of the three quantities are less than $1.5\times10^{-2}$, $1\times10^{-3}$, and $2\times10^{-15}$ for $N=500$\,(red bars), respectively.
These results indicate that the QW PIC with a circuit depth of 12 can be well calibrated by using several hundred training data samples.
We note that our calibration method requires only about a minute for the entire training and validation processes of the network on a laptop equipped with an AMD Ryzen 7 5800H CPU.
The training and validation processes are performed using the PyTorch package for Python program\,\cite{paszke2019}.


\section{\label{sec3}Experimental Setup and Results}
\subsection{Experimental setup}

\begin{figure*}[t]
    \centering
    \includegraphics[width=\linewidth]{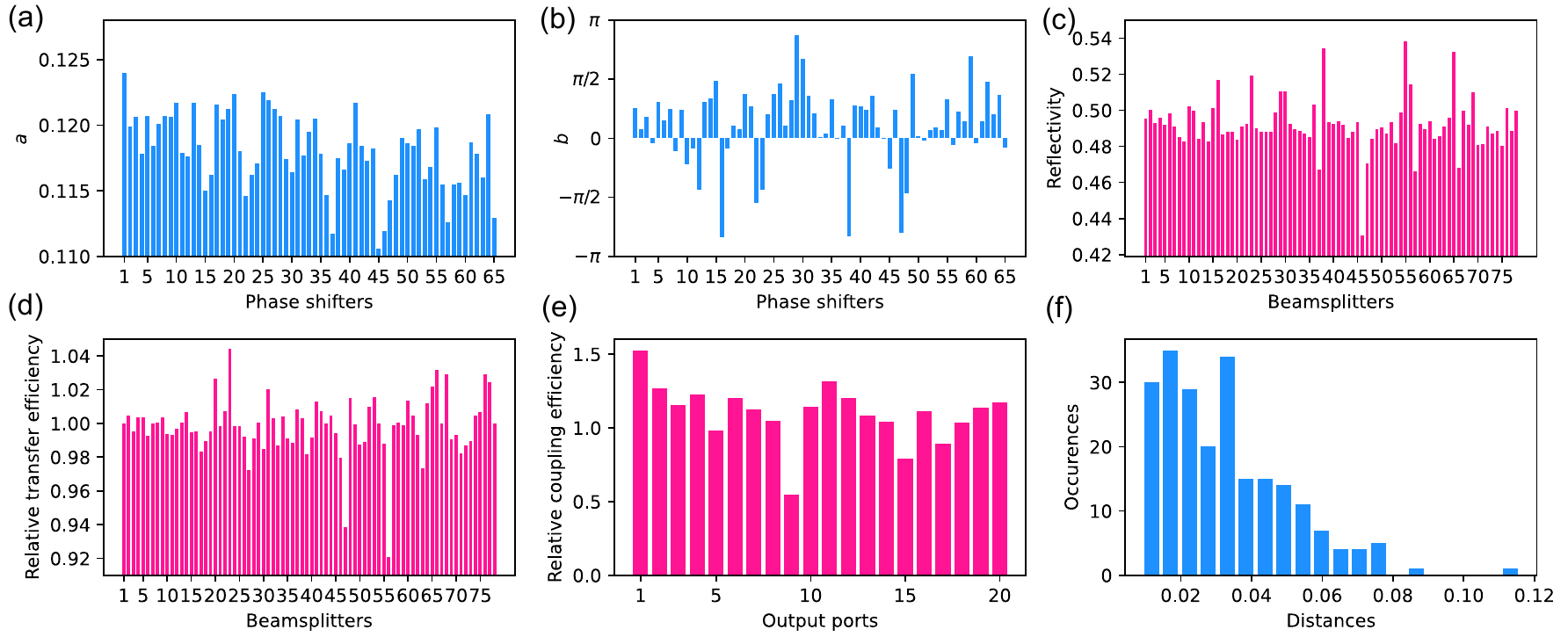}
    \caption{\label{fig:exp_res} 
    \textbf{Experimental calibration results.}
    (a) and (b) depict the predicted values of the parameters $\bm{a}$ and $\bm{b}$ in the phase-current relation for each phase shifter, respectively.
    (c) Predicted values of the reflectivity $\sin^2\bm{\theta}$ for individual on-chip beam splitter.
    (d) Predicted relative transfer efficiency $\bm{\alpha}$ between the beam splitters.
    (e) Predicted relative coupling efficiency $\bm{\eta}$ between the output ports.
    (f) Histogram of the measured values of the L1-norm distance $d(\bm{\lambda}^{\ast})$, which represent the difference between the test dataset's output light probability distributions $\{\bm{P}^{\text{test}}_{\text{exp}}\}_n$ and the trained network's predictions $\{\bm{P}_{\text{ML}}(\bm{\lambda}^{\ast})\}_n$.
    }
\end{figure*}

In our experiment, the conceptual diagram of the QW PIC depicted in Fig.\,\ref{fig:chip} is implemented through silicon waveguide circuits fabricated using the CMOS-compatible silicon photonics process\,\cite{Feng2022}, as illustrated in Fig.\,\ref{fig.setup}.
The $6.8\,\text{mm}\times3.4\,\text{mm}$ chip comprises an array of 66 individually controlled thermo-optic phase shifters with 78 waveguide beam splitters to form the cascade of the interference network of a 12-step on-chip QW.
The controlled phase shifter is implemented by electrically heating a resistor to change the refractive index of the waveguide to deform the optical path length\,\cite{Harris2014}.
Then, the phase shifter has a linear phase-power relation, and its phase change $[0, 2\pi]$ can be tuned by changing the imposed current according to the nonlinear phase-current relation in Eq.\,(\ref{eq:5}).
To program all these phase shifters, we have developed a multichannel electronic control system.
The beam splitter elements are realized using multimode interference couplers.
On-chip grating couplers\,\cite{Son2018} are used for fiber-chip coupling, and the coupling loss is 4.5$\,$dB per facet.
A continuous-wave laser (TLX1, Thorlabs) operating at a wavelength of 1550\,nm and with a power output of 5\,mW is coupled into the QW PIC through a single-mode optical fiber.
The input fiber is precisely aligned with the lower path of the first waveguide beam splitter using a three-axis motorized positioning stage (KWL06050, Suruga), and the polarization of the input laser is fine-tuned using the three-paddle fiber polarization controller (FPC030, Thorlabs).
We simultaneously measure the light intensity from all output ports by coupling them to an array of optical fibers and subsequently directing them to Germanium-amplified detectors (PDA30B2, Thorlabs).
For each output port, we detect the light intensity five times and store the average value.
Consequently, the measured intensity distributions are normalized to the experimental probability distribution $\bm{P}_{\text{exp}}$.
In the experiment, the QW PIC is temperature controlled at 22$^{\circ}$C.

\subsection{Experimental results}

Herein, the calibration of the QW PIC is implemented and tested by training the custom network directly on actual experimental datasets $\{\bm{P}_{\text{exp}}\}_n$.
In order to generate the experimental training and test datasets, we randomly generate 1500 sets of current distributions $\{\bm{I}\}_n$ ranging from 0\,mA to 7\,mA imposed on the phase shifters on our silicon PIC.
We obtain 1500 sets of experimental probability distributions $\{\bm{P}_{\text{exp}}\}_n$ after normalizing the measured light intensity over all output ports.
The experimental datasets are divided into two parts according to an 85:15 ratio, where 1275 sets of datasets are used to train the custom network, and the remaining 225 sets of datasets are used to test the trained network.
With these experimental training sets, the parameters $\bm{\lambda}=\{\bm{a},\bm{b},\bm{\theta},\bm{\alpha},\bm{\eta}\}$ of the custom network in Fig.\,\ref{fig:scheme} are then pretrained.

The trained network can characterize the phase-current relation, assumed to be of the form $\phi = aI^2 + b$, the reflectivity parameter $\bm{\theta}$ and the relative transfer efficiency $\bm{\alpha}$ of BSs, as well as the relative output coupling efficiency $\bm{\eta}$ between the output ports.
Here, we provide the relative transfer efficiency between all BSs and the relative coupling efficiency between all output ports because we utilize the normalized probability distributions to calibrate the QW PIC.
Even so, we need only to measure the absolute transfer efficiency for one BS and the absolute coupling efficiency for one output port.
The efficiencies of all the remaining BSs and output ports can then be directly calculated.

The characterization results for 65 phase shifters, 78 beam splitters and 20 output ports are shown in Figs.\,\ref{fig:exp_res}(a)-\ref{fig:exp_res}(e).
Note that we only experimentally detect 20 output ports in the middle, neglecting two output ports at the top and bottom positions, as the output light intensities at these four ports are not affected by the phase shifter settings.
The phase shifter in the upper right corner of the QW PIC is also omitted because it has no impact on the output light intensity.
Due to the influence of the chip fabrication errors, the coefficients $\bm{a}$ and $\bm{b}$ in Figs.\,\ref{fig:exp_res}(a) and \ref{fig:exp_res}(b) for each phase shifter are often not the same, and the reflectivity of some beam splitters in Fig.\,\ref{fig:exp_res}(c) deviates from the ideal value of 0.5.

\begin{figure*}[t]
    \centering
    \includegraphics[width=\linewidth]{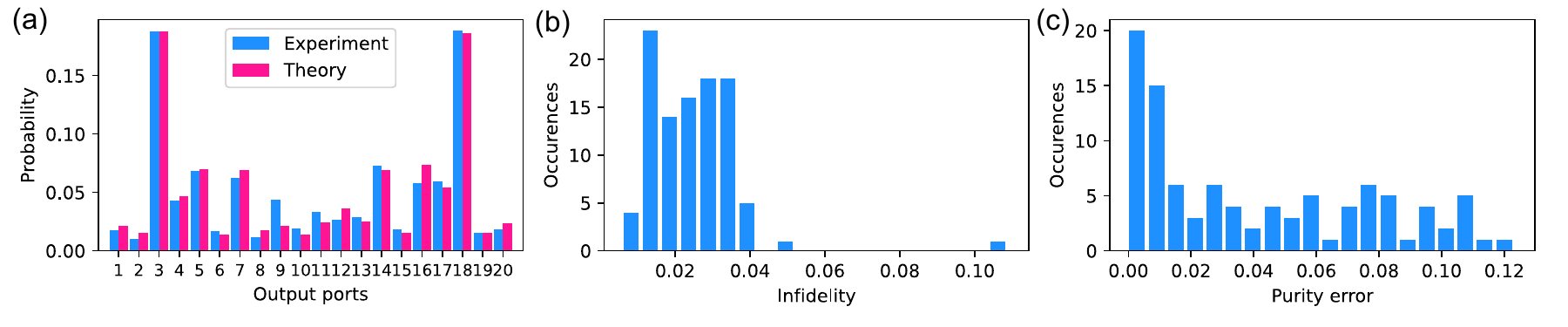}
    \caption{\textbf{Validation of the QW PIC.} (a) The measured probability distribution $\bm{P}^{\text{QW}}_{\text{exp}}$ (blue bars) for 12-step Hadamard QWs.
    The red bars mean the distribution $\bm{P}_{\text{ML}}(\bm{\lambda}^{\ast})$ predicted by the trained network.
    (b) and (c) display histograms of the measured values of infidelity and purity error for 100 output quantum states after nine-step QW dynamics, respectively.
    \label{fig:QW}}
\end{figure*}

After completing the calibration of the photonic chip using the training datasets, we utilize the experimental test datasets to further verify the accuracy of the calibration.
We apply the remaining 225 sets of distributions of random currents, and compare the experimentally measured distributions $\{\bm{P}^{\text{test}}_{\text{exp}}\}_n$ with the predicted probability distributions $\{\bm{P}_{\text{ML}}(\bm{\lambda}^{\ast})\}_n$ of the trained network with the optimal hyperparameters $\bm{\lambda}^{\ast}$.
The difference between the two distributions for each test sample $n$ is defined by L1-norm distance: $d(\bm{\lambda}^{\ast}) = \frac{1}{2}|\bm{P{_\text{ML}}}(\bm{\lambda}^{\ast}) - \bm{P}^{\text{test}}_{\text{exp}}|$.
We observe that the distance values for $85\%$ of the test samples are less than 0.05, as shown in Fig.\,\ref{fig:exp_res}(f).
The average distance for all test samples is $\bar{d}=0.033\pm 0.018$, indicating that our calibration results for the phase shifters and beam splitters are in good agreement with the actual physical process.
The slight deviation suggests the presence of additional chip imperfections that are not considered by our custom network, such as the nondominant thermal crosstalk effect\,\cite{Ceccarelli2020}.
The error bar of $\bar{d}$ represents one standard deviation.

After verifying the performance on numerous random validation samples using the calibrated photonic chip, we proceed to implement the conventional 1D Hadamard QWs, where the coin operators are fixed as the Hadamard gate and independent of time and position.
We inject 1550-nm light into the lower path of the first waveguide beam splitter followed by setting the phase of the first phase shifter to $\pi/2$.
Therefore, we prepare a balanced superposition initial state between two paths with a phase difference of $\pi/2$ through one-step QW dynamics\,\cite{Wang2022}.
The phases of the remaining phase shifters are set to 0.
We measure the light intensity coupled to the optical fiber arrays, and the experimental probability distribution $\bm{P}^{\text{QW}}_{\text{exp}}$ for a 12-step QW is obtained using the calibrated fiber-coupling efficiency for compensation, as depicted in Fig.\,\ref{fig:QW}(a).
The resulting L1-norm distance with respect to the QW distribution predicted by the trained network is $d=0.057$, indicating their good agreement.

Beyond the position distributions, we then perform the quantum state tomography for the calibrated QW PIC with inherent high-dimensional structures\,\cite{wang2024,Titchener2016,Zia2023,Suprano2024}.
To achieve this, in addition to the computational basis $\ket{v}$, extra measurement bases need to be introduced to fully characterize the output quantum state.
We divide the QW PIC into two sections: the first part, with a circuit depth of nine, is used to implement arbitrary nine-step QW dynamics, while the second part, with a circuit depth of 3, serves as a high-dimensional measurement device\,\cite{Innocenti2017,Giordani2019,Zhao2015,Xiaowei2023}. 
For each nine-step QW dynamics, we impose 14 sets of current distributions into the measurement device and measure the output probability distributions.
Here, we perform standard maximum-likelihood reconstruction with the estimator taking the form ${\rho_{\text{m}}=TT^\dagger/\text{Tr}(TT^\dagger)}$, where $T$ is a lower triangular matrix\,\cite{Shang2017}.
The quantum states after the 9-step QW can be reconstructed by maximizing the likelihood of 14 sets of experimental distributions. 
We arbitrarily generate 100 types of QW dynamics, and the histograms of the infidelity and purity error of the reconstructed quantum states are displayed in Figs.\,\ref{fig:QW}(b) and \ref{fig:QW}(c), respectively.
The corresponding average values of the infidelity and purity error are $0.025\pm0.012$ and $0.042\pm0.037$, respectively, indicating that the experimental states align well with the theoretical expectations.

\section{\label{sec4}Conclusion}

In conclusion, we introduce a global calibration method for the integrated photonic circuits.
To achieve this, we construct a network with custom layers to model the physical process of the chip and train it using the gradient-descent-based tool, Adam. 
We experimentally validate our calibration method using a photonic QW chip with a circuit depth of 12.
The trained network can globally calibrate the nonlinear phase-current relations of the thermo-optical phase shifters and the reflectivities of beam splitters on the QW chip, which efficiently bypasses the error accumulation problem associated with the traditional one-by-one calibration method.
Consequently, a low average distance of 0.033 between the measured chip distributions and the network predictions is achieved.

Our calibration method does not require a meticulous design of the decoupling scheme to calibrate phase shifters one by one, making it suitable for photonic chips with a prior physical structure.
Moreover, this method offers high degrees of freedom, allowing any parameters involved in the physical process to be added to the custom network and learned through gradient-descent-based optimization.
For instance, thermal crosstalk parameters between phase shifters could be further characterized by computing the nondiagonal terms in a given matrix phase-current relation.
Thus, our method shows promise as a potent tool with potential applications in large-scale chip calibration and optimal control problems\,\cite{Burgwal2017,Jacques2019,Bandyopadhyay2021,Fyrillas2024}.

\begin{acknowledgments}
J.-H. Zheng and Q.-Q. Wang contributed equally to this paper. This work was supported by Innovation Program for Quantum Science and Technology (Nos.\,2021ZD0301200, 2021ZD0301500, 2021ZD0303200), National Natural Science Foundation of China (Nos.\,12022401, 62075207, 62275240, 11874343, 12104433, 11821404, 12204468), China Postdoctoral Science Foundation (No.\,2024M753083), National Postdoctoral Program for Innovative Talents (No.\,BX20240353), Students' Innovation and Entrepreneurship Foundation of USTC (No.\,SC5290005208), Fundamental Research Funds for the Central Universities (Nos.\,WK2470000030, WK2030000081), CAS Youth Innovation Promotion Association (No.\,2020447), and Research and Development Program of Anhui
Province (No.\,2022b1302007). We thank Hefei Guizhen Chip Technologies Co., Ltd. for collaborating to develop the multi-channel current/voltage source.
\end{acknowledgments}

\appendix

\bibliography{ref}

\end{document}